# Copula-based local dependence between energy, agriculture and metal commodity markets

Claudiu T. Albulescu[a*], Aviral K. Tiwari[b], Qiang Ji[c,d]

[a] Management Department, Politehnica University of Timisoara, Timisoara, Romania
[b] Rajagiri Business School, Rajagiri Valley Campus, Kochi, India
[c] Institutes of Science and Development, Chinese Academy of Sciences
[d] School of Public Policy and Management, University of Chinese Academy of Sciences, Beijing, China

**Abstract:** This paper studies the extreme dependencies between energy, agriculture and metal commodity markets, with a focus on local co-movements, allowing the identification of asymmetries and changing trend in the degree of co-movements. More precisely, starting from a non-parametric mixture copula, we use a novel copula-based local Kendall's tau approach to measure nonlinear local dependence in regions. In all pairs of commodity indexes, we find increased co-movements in extreme situations, a stronger dependence between energy and other commodity markets at lower tails, and a 'V-type' local dependence for the energy-metal pairs. The three-dimensional Kendall's tau plot for upper tails in quantiles shows asymmetric co-movements in the energy-metal pairs, which tend to become negative at peak returns. Therefore, we show that the energy market can offer diversification solutions for risk management in the case of extreme bull market events.

.

---

[*] Corresponding author. E-mail address: claudiu.albulescu@upt.ro, claudiual@yahoo.com.

# 1. Introduction

Energy prices' co-movements with other commodities' prices have generated a huge body of literature given their implications for the economy. We add to this literature by studying the extreme co-movements between energy, agriculture and metal commodity markets[1], using sub-indexes of Rogers International Commodity Index. Different from previous studies, we focus on commodity market general indexes and not on individual commodity prices, to better capture the extreme dependence generated by the substitution effect between specific commodities, amplified by recent innovations and environmental concerns. In doing so, we first identify non-parametric mixture copulas that better fit the pair-wise combinations compared with individual copulas. As a novelty, starting from the best-fitted copula, we use a copula-based local Kendall's tau approach to measure nonlinear local dependence in regions between commodity markets. As far as we know, this is the first paper addressing this issue and different from previous findings, we report asymmetric co-movements between energy and metal prices, which tend to become negative at peak returns.

The price co-movements of different categories of commodities belonging to the main commodity markets (e.g., energy, agriculture and metal) have been analysed from various angles and received special attention following oil price shocks and after the recent food price crisis from 2006 to 2008. These co-movements are explained by economic channels as production and transportation costs, increasing demand for biofuels and use of renewables, substitution effect between commodities generated by disruptive technologies, which have generated specific mechanism for price transmission (Jiang et al., 2018). The financial channels are at work as well, given the existence of higher returns of commodity prices compared with classic financial assets (Brooks and Prokopczuk, 2013) and opportunities they provide for portfolio diversification and risk hedging (Doran and Ronn, 2008; Rafik and Bloch, 2016).

---

[1] According to the Commodity Market Outlook (October 2019), the energy, agriculture and metal commodity markets represent the main categories of commodities (World Bank, 2019).

Therefore, recent empirical analyses focus on price co-movements (e.g., De Nicola et al., 2016; Lucotte, 2016; Pal and Mitra, 2017; Tiwari et al., 2018; Su et al., 2019), or investigate the price volatility spillovers between energy, agriculture and metal commodity prices (e.g. Aguilera and Radetzki 2017; Behmiri and Manera, 2015; Fasanya and Akinbowale, 2019; Ji et al., 2018a,b, 2019; Luo and Ji, 2018; Mensi et al., 2014). However, only few papers investigate the extreme co-movements and resort to copula analyses, using various families of time-invariant and time-varying individual copula functions (Reboredo, 2012, 2013; Reboredo and Ugolini, 2016; Ji et al., 2018a; Jiang et al, 2018; Yahya et al., 2019). However, these papers analyse the extreme dependence between specific commodity prices (i.e. oil and gold, or oil and biofuels), without considering the overall commodity markets. Moreover, copula functions used in previous papers model global dependence structure in energy markets but are unable to measure nonlinear local dependence in regions. In other words, copulas might cover useful information about the changing trends of the degree of commodity markets co-movement and about the asymmetric co-movement in bear and bull markets.

To overcome this limitation, we first use a novel copula-based local dependence framework recently proposed by Huang et al. (2018), which nests the concepts of global dependence and tail dependence. More precisely, we apply the local Kendall's tau based on the mixture copula to identify co-movements and extreme dependencies between agriculture, metal and energy commodity markets that are invisible in a global framework.[2] The investigation of extreme dependencies in regions have multiple advantages compared with the analysis of global dependencies. On the one hand, investigating the local dependencies allows to see the changing trend of the degree of co-movement between two series. Whereas time-varying copulas allow for analysing the time dynamics of the dependence, they do not identify asymmetries between bull and bear markets. Huang's et al. (2018) approach addresses this issue and proposes four classes of local dependence measures along both the main diagonal and the minor diagonal respectively. On the other hand, this approach provides a more detailed dependence information between commodity markets, as it uncovers the relationship between copula functions and the rank-based local dependence measures. Huang's et al. (2018) approach is implemented in three steps. We first filter the data to remove the index returns serial correlation and conditional heteroskedasticity. To this end, like Huang et al. (2019), we resort to an ARMA(p,q)- GARH(m,n) model. Second, we investigate the global dependencies using individual copulas accounting for asymmetric and heavy-tail dependencies (i.e., Gumbel, Clayton, rotated Gumbel and rotated Clayton copulas), as well as several two-component mixtures.[3] In doing so, we apply the Zimmer's (2012) approach for mixture copulas and we find that the best-fitted copula is a mixture between Gumbel and rotated Gumbel copulas (180 degrees). We also perform a rolling window analysis of the best mixture copula to see how extreme dependencies evolve over time. Third, starting from this copula function, we compute the empirical local Kendall's tau and we draw the corresponding theoretical and empirical local dependence surfaces in different regions (upper-upper, lower-lower, upper-lower and lower-upper tails), conducting a series of 5,000 Monte Carlo simulations. As in Huang et al. (2018), we restrict the quantiles of our series in the interval [0.05, 0.95] to ensure enough observations to calculate the empirical local Kendall's tau. Finally, we rank the dependence structure of commodity markets and we underline the advantages of using local dependence measures for investigating the extreme co-movements in commodity markets.

---

[2] Kendall's tau represents a rank dependence coefficient used to measure dependencies of financial series and can be expressed via copula functions (see for example Chollete et al. (2009) and Rodriguez (2007)).

We also contribute to the exiting literature by performing an analysis of extreme dependencies and co-movements between general commodity markets, rather than between prices of individual commodities. This is important to capture the global dynamics of commodity markets. Indeed, inside each category of commodities (i.e. energy, agriculture or metal), the price behaviour of each element might record opposite trends in the presence of a substitution effect (e.g. the price of natural gas and coal, or the price of copper and aluminium), generating thus a distortion in the analysis of price transmissions and co-movements of different categories of commodities. Another reason that recommend the use of composite commodity market indexes is represented by the focus of institutional investors on a broad-based international vehicle, which makes the investigation of extreme dependencies particularly appealing. For this purpose, we use three composite Rogers International Commodity Indexes, namely Rogers International Commodity Index Energy (RICIE), Rogers International Commodity Index Agriculture (RICIA) and Rogers International Commodity Index Metals (RICIM). The advantages of using these indexes consist thus in their ability to characterise the overall commodity markets by considering the prices of 38 commodity futures contracts, with different weights (i.e., the crude oil futures prices represent 37.50% from RICIE, while the natural gas prices represent 15%; gold represents 19.92% from RICIM, while aluminium represents 15.93%; corn prices represent 13.61% from RICIA whereas cotton represents 12.03%).[4] At the same time, the composition of these indexes is considered more stable compared to the composition of other commodity indexes, which increases the transparency for making investment decisions. Finally, yet importantly, these indexes are constructed based on futures rather than spot prices.[5] As far as we know, Kang et al. (2017) and Liu et al. (2019) are the only paper that addresses the dynamic spillover between commodity markets, resorting to futures contract prices.

---

[3] Huang's et al. (2018) methodology can be applied to different families of symmetric and asymmetric copula functions. However, most previous studies underline the advantages of asymmetric copulas, which determined Huang et al. (2018) to estimate the theoretical local dependence in region for the Gumbel, Clayton, and their rotated versions. At the same time, the mixture of these asymmetric copulas in upper and lower tails, is considered to capture more flexible dependence structure compared with individual copulas (see Hu, 2006).

---

[4] http://www.rogersrawmaterials.com/documents/RICIHndbk_01.31.19.pdf
[5] Investigating the co-movement and extreme dependencies between futures prices have multiple advantages, as it allows for the mitigation of stale quote and non-synchronous problems existing on the spot markets, while the noise in the futures prices is constant in average (for a detailed discussion, please refer to Albulescu et al., 2017).

The last contribution to the literature is represented by a portfolio analysis considering the pairs of commodity indexes. Starting from the time-varying dynamic conditional correlation (DCC) model of Engle (2002) we compute hedge ratios following Kroner and Sultan (1993). Afterwards, we use conditional volatilities to construct optimal portfolio weights, resorting to the Kroner and Ng's (1998) approach.

To preview our findings, we show that co-movements increase in extreme situations for all pairs of commodity indexes, a result in agreement with previous findings reported in the literature. However, in contrast to other studies, we discover a stronger dependence at lower tails for the energy-pairs of indexes. In addition, for the agriculture-metal pair we report a 'V-type' local dependence. Furthermore, the three-dimensional Kendall's tau plot for upper tails in quantiles shows asymmetric co-movements for the energy-metal pair, which becomes negative at peak returns. Finally, we document the existence of hedging and portfolio diversification opportunities between energy, agriculture and metal commodity markets.

The rest of the paper is as follows: Section 2 presents the literature review on commodity prices co-movements; Section 3 addresses the methodology; Section 4 presents the data and the marginal model results; Section 5 describes the copula results; and Section 6 shows the general and copula-based local dependence plots. The last section concludes.

**2. Literature review**

The literature addressing the co-movements and dependencies between commodity markets usually focuses on individual commodity prices and explains the co-movements by exploiting both economic and financial channels.

A first strand of the literature focuses on the relationship between energy and agriculture commodities, addressing the price co-movements and volatility spillovers. In terms of price co-movements, early studies use linear cointegration techniques (Avalos, 2014; Baumeister and Kilian, 2014; Nazlioglu and Soytas, 2012; Saghaian, 2010; Rezitis, 2015; Liu et al., 2017) or multivariate linear regressions (Hassouneh et al., 2012) and document the existence of long-run co-movements or volatility spillovers (Du et al., 2011; Fasanya and Akinbowale, 2019; Ji and Fan, 2012; Serra, 2011; Nazioglu et al., 2013; Mensi et al., 2014; Zhang and Qu, 2015). Studies that are more recent focus on the non-linearity characterising this relationship (Chen et al., 2010; De Nicola et al., 2016; Lucotte, 2016; Natanelov et al., 2011; Pal and Mitra, 2017; Su et al., 2019) and document increased co-movements between energy and agricultural commodity prices, following the recent food crisis and the rise of environmental concerns. In contrast, Fowowe (2016) uses cointegration tests with structural breaks and nonlinear causality tests and shows that agriculture commodity prices do not respond to oil price shocks in South Africa.

A second bulk of the literature addresses the co-movement and volatility spillover between energy and metal markets (Aguilera and Radetzki 2017; Behmiri and Manera, 2015; Bildirici and Turkmen. 2015; Choi and Hammoudeh, 2010; Ewing and Malik, 2013; Hammoudeh and Yuan, 2008; Ji et al., 2018b. For example, Choi and Hammoudeh (2010) investigate the volatility transmission between oil and industrial commodities using a regime-switching model, while Bildirici and Turkmen (2015) analyse the cointegration and causality relationship among oil and precious metals using a nonlinear ARDL cointegration framework and nonlinear causality tests. Whereas Aguilera and Radetzki (2017) investigate the synchronisation of oil and gold prices, Behmiri and Manera (2015) underline the role of oil price shocks to explain volatility in metal prices.

We notice that the literature assessing the co-movements and volatility spillovers between commodity prices is quite extensive. Nevertheless, only few papers resort to copula functions to address the extreme co-movements and dependencies on commodity markets. A first paper in this line is the work by Reboredo (2012), which investigates the dependencies between oil and food prices using conditional and time-varying copulas and report the absence of extreme dependencies. With a focus on extreme dependencies between the energy and metal markets, Reboredo (2013) explores the possibility that gold might represents a hedge against oil price volatility. Reboredo and Ugolini (2016) underline the role of oil price shocks to explain volatility in metal prices. Adopting a time-varying copula with a switching dependence, the study by Ji et al. (2018a) shows that agricultural commodities are more sensitive to oil price shocks than gas price shocks. In the same line, Liu et al. (2019) resort to Markov-switching GRG mixture copulas to study the dependencies between oil and agriculture futures prices in both non-extreme and extreme conditions. For the majority of considered agriculture commodities, the authors report positive extreme and non-extreme co-movements with oil futures prices. Jiang et al. (2018) and Yahya et al. (2019) combine time-frequency decomposition of commodity price series with copula functions to study the extreme dependencies between energy, agriculture and metal commodity prices. Both papers document an increased tail dependence during and after the recent global crisis.

We add to this narrow strand of the literature and, in contrast to these studies that use individual time-invariant and time-varying copulas, we resort to mixture copulas to explain the extreme dependencies of commodity markets more clearly, considering asymmetries in both upper and lower tails. Further, to gain additional information about the extreme co-movements in the upper and lower tails, we use a local dependence measure based on the Kendall's tau, which combines the global and tail dependence.

## 3. Methodology

### *3.1. Marginal model*

We use an ARMA(p,q)-GARCH(m,n) model with skewed t-distribution to remove the serial correlation and conditional heteroskedasticity in all index return series. The conditional mean of the log-return series follows an ARMA(p,q) process:

$$r_t = \mu + \sum_{i=1}^{p} \phi ar_i\, r_{t-i} + \sum_{j=1}^{q} ma_j\, \varepsilon_{t-j} + \varepsilon_t, \tag{1}$$

where $\varepsilon_t$ is a white noise which follows the Student-t distribution, such as:

$$\sqrt{\frac{\vartheta}{\sigma_t^2(\vartheta-2)}}\, \varepsilon_t \sim iid\ t_\vartheta, \tag{2}$$

where $\vartheta$ are the degrees of freedom and $\sigma_t^2$ is the conditional variance of $\varepsilon_t$, given by: $\sigma_t^2 = \omega +$

$$\sum_{i=1}^{p} \alpha_i \varepsilon_{t-p}^2 + \sum_{j=1}^{q} \beta_j \sigma_{t-j}^2, \tag{3}$$

where $\omega$ is a constant, $\varepsilon_{t-p}$ is the ARCH component, while $\sigma_{t-j}^2$ is the GARCH component. The number of lags (p, q) is selected according to the Akaike Information Criteria (AIC).

### 3.2. Copula functions

Given the data properties that exhibit asymmetries and heavy tails, we use the Gumbel copula which displays upper-tail dependence, the Clayton copula which displays lower-tail dependence, their rotated versions and their mixture.

The bi-variate Gumbel copula ($C_G$) is expressed as:

$$C_G(u, v; \alpha) = \exp\left(-[(-\log(u))^\alpha + (-\log(v))^\alpha]^{1/\alpha}\right), \tag{4}$$

where $\alpha \in [1, +\infty]$. When $\alpha \to +\infty$, the variables exhibit more dependence.

The rotated Gumbel copula at 180 degrees ($C_{rG}$) is:

$$C_{rG}(u, v; \alpha) = u + v - 1 + C_G(1 - u, 1 - v; \alpha), \tag{5}$$

where $\alpha \in [1, +\infty]$. In this case also, if $\alpha \to 1$, the two variables are independent.

The Clayton copula ($C_{CL}$) shows asymmetry, as the degree of dependence in the lower tail is higher than in the upper tail:

$$C_{CL}(u, v; \theta) = [\max(u^{-\theta} + v^{-\theta} - 1, 0)]^{-1/\theta}, \tag{6}$$

where $\theta \in [-1, 0] \backslash \{0\}$ and larger values of $\theta$ indicate strong dependence.

The rotated Clayton copula ($C_{rCL}$) only shows lower tail dependence. Similar to $C_{CL}$, larger values of $\theta$ indicate strong dependence:

$$C_{rCL}(u, v; \theta) = u + v - 1 + C_{CL}(1 - u, 1 - v; \theta). \tag{7}$$

Building upon Zimmer (2012) and Huang et al. (2018), we define the mixture of four individual copulas as follows:

$$C_{mix1}(\cdot) = \varphi * C_G(\cdot) + (1-\varphi) * C_{CL}(\cdot), \quad (8)$$

$$C_{mix2}(\cdot) = \varphi * C_{rG}(\cdot) + (1-\varphi) * C_{rCL}(\cdot), \quad (9)$$

$$C_{mix3}(\cdot) = \varphi * C_G(\cdot) + (1-\varphi) * C_{rG}(\cdot), \quad (10)$$

$$C_{mix4}(\cdot) = \varphi * C_{CL}(\cdot) + (1-\varphi) * C_{rCL}(\cdot), \quad (11)$$

where $\varphi \in (0,1)$ is an estimable parameter indicating the proportional contribution of the first copula in the mixture.

### 3.3. General Chi-plots for dependence identification

The dependencies pattern between two series can by identified by using Chi-plots, which are estimated based on the joint distribution ($H$) of copulas and show the shape of this joint distribution. Following Fisher and Switzer (1985), we generate a scatter plot for each pair of indexes ($\lambda_i, \chi_i$) based on:

$$\lambda_i = 4S_i \max\left\{\left(F_i - \frac{1}{2}\right)^2 \left(G_i - \frac{1}{2}\right)^2\right\} \text{ and} \quad (12)$$

$$\chi_i = \frac{H_i - F_i G_i}{\{F_i(1-F_i)G_i(1-G_i)\}^{0.5}}, \quad (13)$$

where $\lambda_i$ is a measure of distance from the center of the data set of the data point ($X_k, Y_k$), $H_i = \sum_{k \neq i} \frac{I(X_k \leq X_i; Y_k \leq Y_i)}{n-1}$, $F_i = \sum_{k \neq i} \frac{I(X_k \leq X_i)}{n-1}$, $G_i = \sum_{k \neq i} \frac{I(Y_k \leq Y_i)}{n-1}$, and $S_i = sign\left\{\left(F_i - \frac{1}{2}\right)\left(G_i - \frac{1}{2}\right)\right\}$.

### 3.4. Copula-based local Kendall's tau plots

The copula-based formula for the global Kendall's tau (Schweizer and Wolf, 1981) for two variables $X$ and $Y$ is:

$$\tau(X,Y) = 4\int_0^1 \int_0^1 C(u,v)dC(u,v) - 1. \quad (14)$$

For two variables $X$ and $Y$ with $p$ and $q$ their quantiles, the novel tail dependence measures on local Kendall's tau developed by Huang et al. (2018) are:

$$\lambda_{UU}^{Kendall} = \lim_{p \to 1} \lambda_{UU}^{Kendall}(X,Y;p,p), \quad (15)$$

$$\lambda_{UL}^{Kendall} = \lim_{p \to 1} \lambda_{UL}^{Kendall}(X, Y; p, 1-p), \qquad (16)$$

$$\lambda_{LU}^{Kendall} = \lim_{p \to 0} \lambda_{LU}^{Kendall}(X, Y; p, 1-p), \qquad (17)$$

$$\lambda_{LL}^{Kendall} = \lim_{p \to 0} \lambda_{UU}^{Kendall}(X, Y; p, p), \qquad (18)$$

where $\lambda_{UU}^{Kendall}$, $\lambda_{UL}^{Kendall}$, $\lambda_{LU}^{Kendall}$ and $\lambda_{LL}^{Kendall}$ are the upper-upper, upper-lower, lower-upper and lower-lower tail dependence measures based on the local Kendall's tau.

## 4. Data and filtering results

### *4.1. Data sample*

The Rogers International Commodity Indexes represents commodity composite futures price indexes reflecting the price dynamics and expectations for energy, agriculture and metal commodity markets. According to the RICI Handbook (Rogers International Commodity Index, 2019), for the futures contracts, the index rolls over three days and it is rebalanced monthly. RICI includes 38 commodities futures contracts, out of each 6 for the energy index RICIE, 22 for the agriculture commodity index RICIA, and 10 for the metal index RICIM.[6]

Our daily data has been obtained from the Quandl database and covers the period 03:01:2005 to 01:08:2018. Table 1 summarises general statistics from log returns of three indexes (dlnRICIA, dlnRICIE and dlnRICIM). The average returns are positive but close to zero, except for RICIM index, while a higher volatility is observed for RICIE. The high kurtosis shows the presence of extreme values in our series, especially in the case of the energy index. The Jarque–Bera for the normality of the unconditional distribution strongly reject the normality for all three series. The unit root tests of ADF, PP and KPSS show that all three series are stationary.

[Insert Table 1 about here]

---

[6] More details about the weights of each commodity inside the market index and about futures contracts can be found at: http://www.rogersrawmaterials.com/documents/RICIHndbk_01.31.19.pdf.

Figures 1(a)-(c) show the indexes and the dynamics of their log returns. It seems that RICIA and RICIM follows similar patterns while RICIE does not exhibit the upward trend enregistered by the other two indexes in 2011. During the last period of 2014, the RICIA and RICIE values decrease, while the RICIM index shows a slight increase after 2016.

[Insert Figure 1 about here]

*4.2. Marginal model results*

Several ARMA(p,q)-GARH(m,n) marginal models were tested. For all our indexes, the best filtering measure is an ARMA (1,0)-GARCH (2,1) model. The results reported in Table 2 show that the autocorrelation and the conditional heteroskedasticity of all three series are removed according to Ljung–Box and ARCH statistics. Our evidence indicates that the AR term is not significant for RICIA.

[Insert Table 2 about here]

## 5. Copula results
*5.1. Individual and mixture copulas*

In what follows, we present a comparison between the results of the individual (Table 3) and mixture copulas proposed in Eqs. 8-11 (Table 4). According to the information criteria (LR, AIC, BIC), in the case of individual copulas, the rotated Gumbel performs better for all commodity pairs, showing asymmetric co-movements which are more important for the lower tails. For all the commodities pairs (RICIA-RICIM, RICIA-RICIE and RICIM-RICIE), our findings show that mixture copulas perform better than the individual copulas, according to all information criteria we use.

[Insert Table 3 about here]

We observe in Table 4 that the best copula mixture model for all commodities pairs is $C_{mix3}(\cdot)$, namely a combination between the Gumbel and rotated Gumbel copulas. For the RICIA-RICIM and RICIM-RICIE pairs, the weight of the Gumbel copula ($\varphi$) is 24.8% and 29.1% respectively, while the RICIA-RICIE pair demonstrate the weight of the Gumbel copula at 40.3%. This evidence of asymmetry (especially for the RICIA-RICIM and RICIM-RICIE co-movements), shows that the rotated Gumbel copula has the largest weight in the mixture (i.e. 75.2% for the RICIA-RICIM pair). Therefore, we notice an increased dependence in both upper- and lower-tails, with the co-movements in the lower-tails being more important (the co-movements are higher in crisis periods).

[Insert Table 4 about here]

For the RICIM-RICIE pair, our copula findings are different from those reported by Reboredo and Ugolini (2016), who notice that spillovers from upward oil price movements to metal prices are larger than for downward oil price spillovers. For the RICIA-RICIE pair, extreme downward oil price movements have a negative effect on agricultural commodity prices. Our results validate the production cost channels and contrast with the results of Reboredo (2012), who reports no extreme market dependence between oil and food prices and shows a neutrality of agricultural commodity markets, especially before 2008. The evidence of asymmetric dependencies between oil and agriculture prices show the need to consider the asymmetric effect of oil prices on the stabilisation of food prices.

*5.2. Best mixture copula rolling analysis*

To gain further insights on commodity markets extreme co-movements, we perform a rolling window analysis (500 days) of the best copula mix ($C_{mix3}(\cdot)$), to observe how dependencies fluctuate over time.[7] To this end, starting from the coefficients' standard errors, we generate confidence intervals at levels of 5%. Figure 2 presents the results for all pairs of indexes.

---

[7] This rolling window roughly correspond to two years trading days. Bai et al. (2018) use a similar rolling window in their analysis and underline the advantages of using lengthy windows in copula analysis.

[Insert Figure 2 about here]

Figure 2(a) shows that with the RICIA-RICIM pair, the extreme dependencies are predominantly explained by the rotated Gumbel copula, which has a higher weight in the mixture copula. Therefore, for the period 2007 to 2017, the co-movements between the agriculture and metal commodity markets are more important in lower tails. A different pattern is noticed starting with 2018, when asymmetric co-movements are generally stronger in upper tails.

In the case of the RICIA-RICIE pair (Figure 2(b)), the extreme dependencies are stronger and better explained by the rotated Gumbel copula, which clearly has a higher weight in the mixture. The exception is for the last period in 2015, when the Gumbel copula becomes dominant. These results show that the period after the Global Financial Crisis is characterised by higher dependencies in the lower tails, while the recent period is characterised by asymmetric co-movements, which increases in the upper tails. Finally, Figure 2(c) shows that from 2007 to 2015, there is no clear dominance of the Gumbel or the rotated Gumbel in the mixture, indicating asymmetric extreme co-movements in both bull and bear markets. Nevertheless, for the last period the extreme dependencies are predominantly explained by the Gumbel copula.

The true parameter values (Figure 2) do not explain, however, the intensity of dependencies over time. Therefore, we proceed to a transformation of true values in Kendall's tau values (Figure 3). As in the previous case, we notice a relative dominance of the rotated Gumbel copula for the sample's first period. Starting with 2016 for all commodity-pairs, we observe a dominance of the Gumbel copula, meaning that the extreme dependencies manifest particularly in the upper tails. If we analyse the dynamics of the mixture copula ($C_{mix3}(\cdot)$) for the RICIA-RICIM pair (Figure 3(a)), we notice an increased dependence between agriculture and metal markets after the Global Financial Crisis (2010 to 2015). A similar result is obtained for the RICIM-RICIE pair of commodities (Figure 3(c)). However, extreme dependence cycles are observed for the RICIA-RICIE pair over the entire analysed period. This evidence shows that in the case of energy-agriculture pair of indexes, the extreme co-movements manifest independently of the economic context. Consequently, the stabilization of energy prices (i.e. oil prices) have an important effect on the stabilization of agriculture and food prices.

[Insert Figure 3 about here]

## 6. General and copula-based local dependence plots

### 6.1. General Chi-plots results

The Chi-plots show that all pairs of commodity indexes exhibit heavy-tail dependence given that most distribution points are plotted beyond the control lines (+/- 0.05). Figure 4 also shows that the extreme dependencies are positive (an increase in one index price is associated with an increase in the price of the corresponding pair), but they are asymmetric. However, the general Chi-plots do not allow details of the dependencies to be seen, namely if there is an asymmetric local dependence pattern or asymmetry in both upper- and lower-tails.

[Insert Figure 4 about here]

### 6.2. Copula-based local Kendall's tau results

To compare the properties of the global and local dependence Kendall's tau based on $C_{mix3}(\cdot)$, we first compute the corresponding theoretical and empirical local dependence surfaces in different regions, performing Monte Carlo simulations for the product copula (with a sample size of 5,000 observations), for each pair of commodities (see Figures A1-A3 (Appendix A)) and show their relationship. For all commodity pairs, the first two plots (Figures A1(a)-A3(a)) indicate that local Kendall's tau based on $C_{mix3}(\cdot)$ copula shows significant asymmetric characteristics along the main diagonal.

Following this, we compare the properties of the global and local dependence co-movements, using copula-based Kendall's tau plots (Figure 5). Starting from Eqs. (15)-(18), we compute the theoretical global and local Kendall's tau, observing no noticeable difference between the global theoretical and empirical Kendall's tau, and show that the global dependence may cover extreme co-movements. Based on the local Kendall's tau for all commodity pairs, we notice increased co-movements in extreme situations, that is, in the lower and upper quantiles. For all commodity pairs, the local dependence in boom markets is obviously smaller than their local dependence in bear markets, exhibiting an asymmetric local dependence pattern. This result is consistent with the findings reported in the earlier literature, underlining the contagion phenomenon in crisis times (Yahya et al., 2019). We notice a stronger dependence at lower tails in the energy-pairs of indexes and a 'V-type' local dependence in the agriculture-metal commodity pair. In contrast with Jiang et al. (2018), we show that tail dependence does change over time. These results show that the energy market is more connected with other commodity markets during periods of financial stress. At the same time, the extreme dependencies between agriculture and metal commodity indexes manifest for both financial boom and downturn periods.

[Insert Figure 5 about here]

Given the existence of a stronger dependence in the lower and upper tails, we construct three-dimensional Kendall's tau plots for these in quantiles for each commodity pair (Figure 6). For the RICIA-RICIM pair, the results show that the extreme dependencies in the lower tails are higher (Figure 6(b)) and, as expected, are symmetric, while the extreme dependencies in the upper tails are asymmetric (Figure 6(a)). This represents an original result which shows the symmetry/asymmetry not only between upper and lower tails, but also inside the upper and lower tails. The economic intuition of this result is simple and might be explained by economic mechanisms. In bear markets when the global demand diminishes, the price of industrial metals decreases. This is equivalent with a reduction in production costs for agriculture commodities. However, in bull markets speculation reasons might explain the asymmetric co-movements between agriculture and metal commodity markets.

[Insert Figure 6 about here]

Similar findings are reported for the RICIA-RICIE pair of indexes, although the co-movements are not as strong. The asymmetric extreme dependence recorded in the upper tails, can be associated with the substitution effect between energy and agriculture commodities, initially induced by the dominance of synthetic fibres, rubber and fertilizers, plastics, which have replaced natural materials. The rise of environmental concerns reversed this trend during last years.

A slightly different situation is seen for the RICIM-RICIE pair of indexes, where the upper tails co-movement are asymmetric and tend to become negative for the peak returns (Figure 6(e)). The negative dependencies can be explained by technological innovations (e.g. cooper cables are replaced by fibre optic lines, resulted from the petrochemical industry. Two implications derive from this evidence. First, the dependencies between energy and metal markets reinforce the impact of energy prices on inflation, given that oil and gold are leading indicators for the general price level (Bildirici and Turkmen, 2015). Second, negative and asymmetric co-movements in the upper tails between energy and metals prices show that precious metals, such as gold, can represent a hedge against inflation for extreme, upper-tail values. Our findings thus complement the results reported by Aye et al. (2016), who also consider gold to be an inflation hedge in the long run.

Overall, these findings confirm the mixture copula results and reinforce the related literature in offering information about a more extensive dependence in quantiles. Our results show that the extreme dependencies of commodity markets are very complex and manifest differently for specific commodity indexes and reveal asymmetric behaviour of commodity prices, especially for the upper tails.

**7. Portfolio analysis**

For the portfolio analysis, like Sadorsky (2012), we use the time-varying DCC model of Engle (2002), with the correlation estimator ($\rho$):

$$\rho_{ij,t} = \frac{q_{ij,t}}{\sqrt{q_{ii,t} q_{jj,t}}}. \tag{19}$$

The summary statistics of dynamic correlations are presented in Table 5 and show no important differences in terms of correlations between our pairs of indexes.

[Insert Table 5 about here]

Starting from the computed conditional volatilities, we construct in a second step hedge ratios using the Kroner and Sultan's (1993) approach, and optimal portfolio weights following Kroner and Ng (1998).

*7.1. Hedge ratios*

The hedge ratio ($\beta$) proposed by Kroner and Sultan (1993) shows how a long position in the asset *i* can be hedged by a short position in the asset *j*:

$$\beta_{ij,t} = \frac{h_{ij,t}}{h_{jj,t}}, \tag{20}$$

where $h_{ij,t}$ is the conditional covariance between assets *i* and *j*.

The results are presented in Table 6 and show that the average hedge ratio between RICIA (long position) and RICIM (short position) is 0.42. This means that 1-dollar long position in RICIA can be hedged for 42 cents with a short position in RICIM. The average hedge ratio between RICIE and RICIA is 0.67, meaning that 1-dollar long position in energy market can be hedged for 67 cents short position in agriculture market.

[Insert Table 6 about here]

*7.2. Portfolio weights*

The Kroner and Ng's (1998) approach is employed to design optimal portfolio weights ($w$). For two assets, $w_{ij,t}$ represents the weight of the asset $i$ in the 1-dollar portfolio of two assets $i$ and $j$ (the weight of the asset $j$ is therefore $1 - w_{ij,t}$):

$$w_{ij,t} = \frac{h_{jj,t} - h_{ij,t}}{h_{ii,t} - 2h_{ij,t} + h_{jj,t}}. \tag{21}$$

Table 7 indicates that the average weight between RICIA and RICIE for example, is 0.84. This means that for 1-dollar portfolio optimization, 84 cents should be invested in agriculture commodity index whereas 16 cents should be invested in energy commodity index. Further, if we consider the RICIM/RICIE pair, for 1-dollar portfolio optimization, 69 cents should be invested in RICIM and 31 cents in RICIE. We can therefore conclude that agriculture and metal commodity markets can offer some portfolio diversification opportunities for the energy commodity market.

[Insert Table 7 about here]

## 8. Conclusions

Using a copula-based local Kendall's tau approach, this study investigates the local dependencies and co-movements between energy, agriculture and metal commodity markets, relying on RICI indexes over the period from 03:01:2005 to 01:08:2018 (daily data). More precisely, we show that a Gumbel and rotated Gumbel copula mixture better fit the pair-wise combinations for all commodity pairs. We therefore posit that dependence structure is asymmetric and exhibits both high- and low-tail dependence, with the low-tail dependence being stronger. We also show that the novel copula-based local Kendall's tau approach offers a deeper understanding of extreme dependencies compared with the global approach.

Specifically, we notice that in the energy-pairs of indexes, a stronger dependence at lower tails exist, while in the agriculture-metal pair we report a 'V-type' local dependence, where the extreme co-movements are higher in both the upper and lower tails. This behaviour is explained by the complementarity between agriculture and metal markets. When the food demand increases, the production process in the agriculture field follows a similar process, which trigger an increase in metal prices. At the same, time if we consider the case of the recent food crisis from 2007 and 2008 when the agriculture commodities' prices increased dramatically, we notice that we recorded similar dynamics in precious metals' prices.

In contrast to previous essays on commodity market co-movements, we show that the upper tail extreme dependencies between energy and metal commodity markets are asymmetric and negative for high returns. This result is explained by the substitution effect existing between energy and metal commodities, amplified lately by the appearance of disruptive technologies implemented in the electricity generation and vehicle industry.

Our findings reveal some complex and extreme dependencies between agriculture, energy, and metal markets and should be of value to those in the field, in particular, financial investors and risk managers. Although during periods of financial stress the energy market does not offer a good option for portfolio diversification, during financial booms, the extreme co-movements between energy and metal commodity markets become negative and highlight the potential for portfolio diversification.

Finally, it is important to know that the extreme co-movements between energy and agriculture markets manifest for the entire time horizon, while the local dependencies between energy and metal markets are mainly recorded in boom and bust periods. The agriculture and metal markets offer portfolio diversification opportunities for the investors in energy futures market.

**Acknowledgements**

This work was supported by a grant of the Romanian National Authority for Scientific Research and Innovation, CNCS–UEFISCDI, project number PN-III-P1-1.1-TE-2016-0142. Supports from the National Natural Science Foundation of China under Grant No. 71974181, No. 71774152, and Youth Innovation Promotion Association of Chinese Academy of Sciences (Grant: Y7X0231505) are equally acknowledged.

Behmiri N.B. and M. Manera. 2015. The role of outliers and oil price shocks on volatility of metal prices. Resources Policy 46: 139–150.

Bildirici, M.E. and C. Turkmen. 2015. Nonlinear causality between oil and precious metals. Resources Policy 46: 202–211.

Brooks, C. and M. Prokopczuk. 2013. The dynamics of commodity prices. Quantitative Finance, 13: 527–542.

Chen, S., Kuo, H. and C. Chen. 2010. Modeling the relationship between the oil price and global food prices. Applied Energy 87: 2517–2525.

Chollete, L., Heinen, A. and A. Valdesogo. 2009. Modeling international financial returns with a multivariate regime-switching copula. Journal of Financial Econometrics 7: 437–480.

Choi, K. and S. Hammoudeh. 2010. Volatility behavior of oil, industrial commodity and stock markets in a regime switching environment. Energy Policy 38: 4388–4399.

De Nicola, F., De Pace, P. and M.A. Hernandez. 2016. Co-movement of major energy, agricultural, and food price returns: A time-series assessment. Energy Economics 57: 28–41.

Doran, J.S. and E.I. Ronn. 2008. Computing the market price of volatility risk in the energy commodity markets. Journal of Banking and Finance 32: 2541–2552.

Du, X., Yu, C.L. and D.J. Hayes. 2011. Speculation and volatility spillover in the crude oil and agricultural commodity markets: A Bayesian analysis. Energy Economics 33: 497–503.

Engle, R.F. 2002. Dynamic conditional correlation: a simple class of multivariate generalized autoregressive conditional heteroskedasticity models. Journal of Business & Economic Statistics 20: 339–350.

# Appendixes

# Appendix A

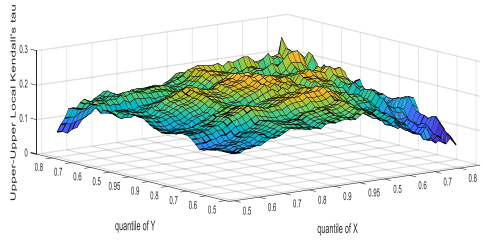 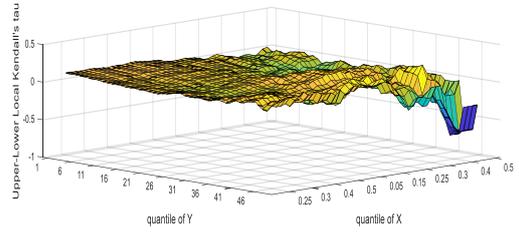
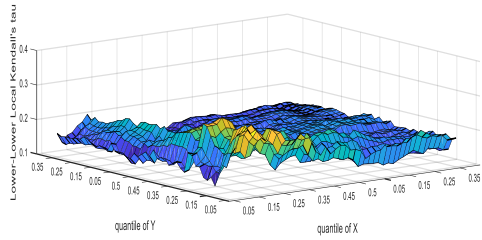 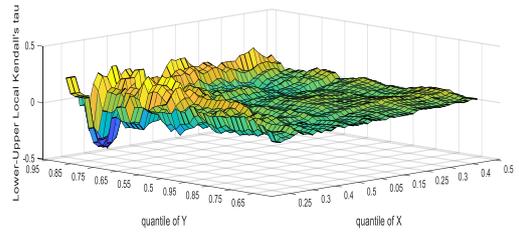

(a) Lower-lower and upper-upper          (b) Upper-lower and lower-Upper

**Figure A1. Local dependence surfaces in different quadrants based on Kendall's tau of Product copula (RICIA-RICIM)**

(Note: The blue surfaces and yellow surfaces are the theoretical and empirical local dependence surfaces)

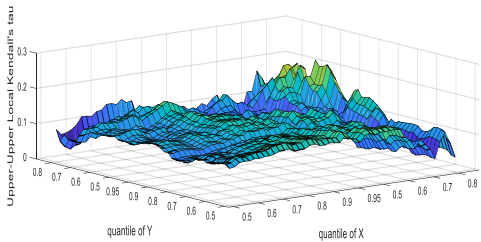 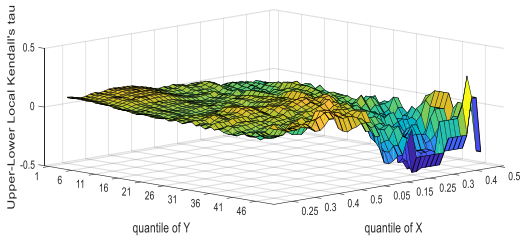
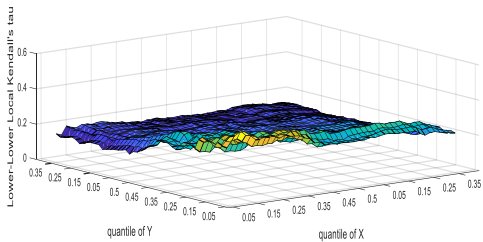 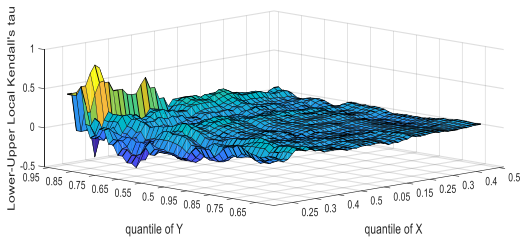

(a) Lower-lower and upper-upper          (b) Upper-lower and lower-Upper

**Figure A2. Local dependence surfaces in different quadrants based on Kendall's tau of Product copula (RICIA-RICIE)**

(Note: The blue surfaces and yellow surfaces are the theoretical and empirical local dependence surfaces)

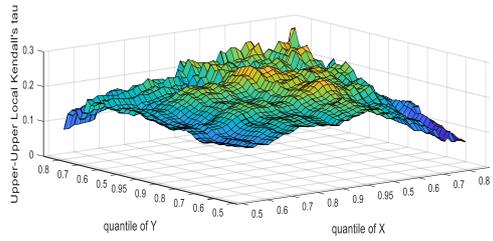 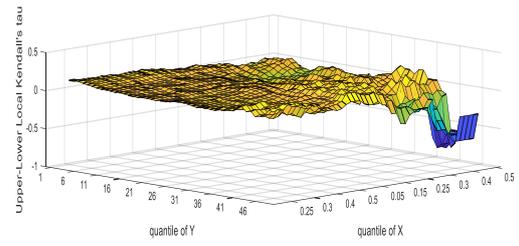
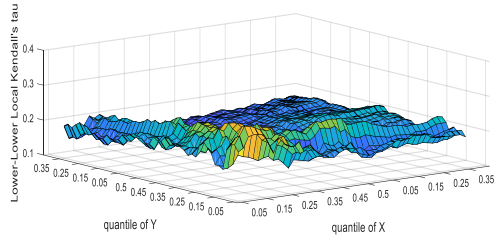 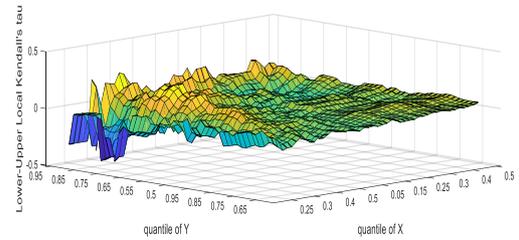

(a) Lower-lower and upper-upper   (b) Upper-lower and lower-Upper

**Figure A3. Local dependence surfaces in different quadrants based on Kendall's tau of Product copula (RICIM-RICIE)**

(Note: The blue surfaces and yellow surfaces are the theoretical and empirical local dependence surfaces)

**Figures**

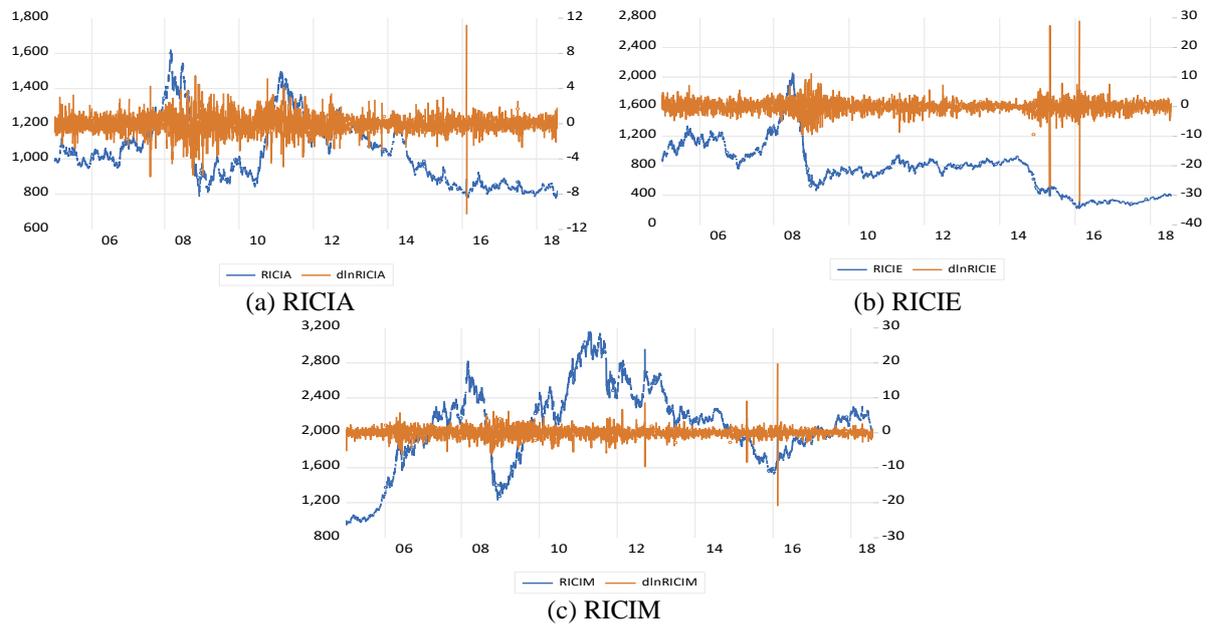

**Figure 1. Commodity price indexes dynamics (log returns and standardised residuals)**

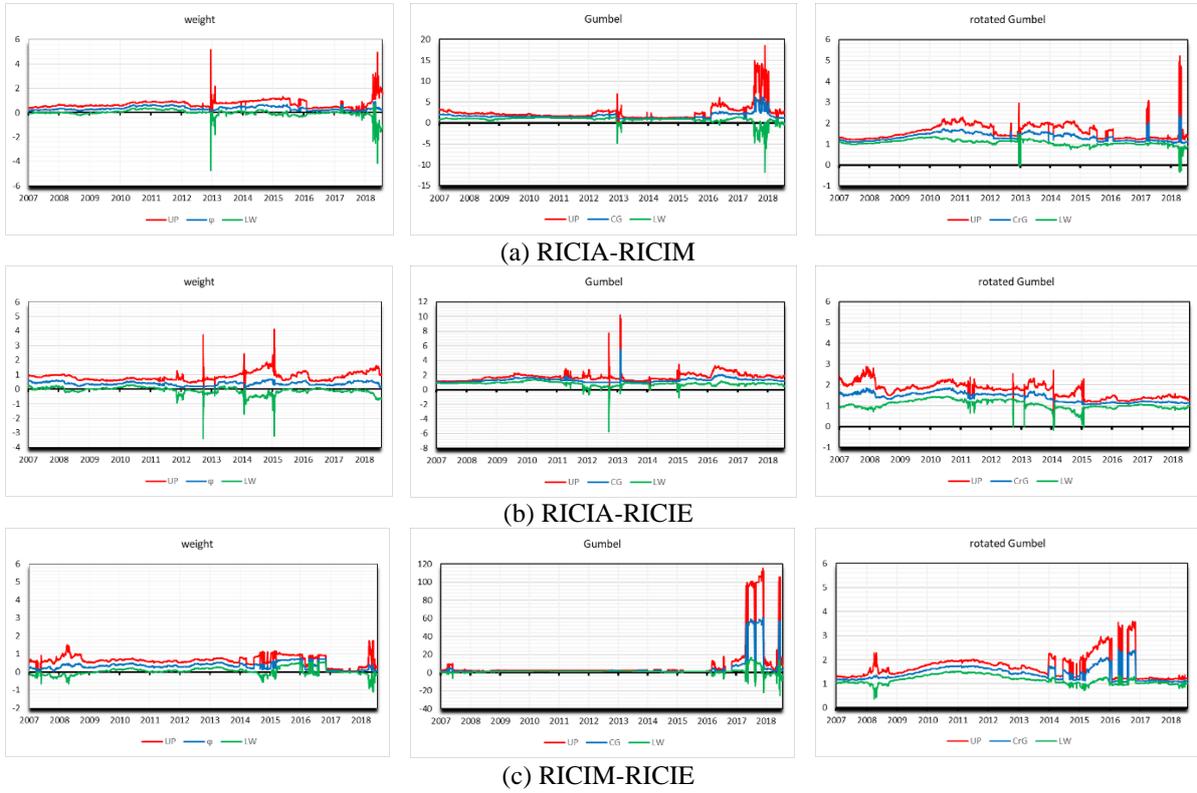

(a) RICIA-RICIM

(b) RICIA-RICIE

(c) RICIM-RICIE

**Figure 2. Rolling window analysis for the best mixture copula**

(Notes: (i) UP (red line) and LW (green line) represent 5% confidence interval limits; (ii) if the blue line lies in the confidence interval, then the parameter is significant.)

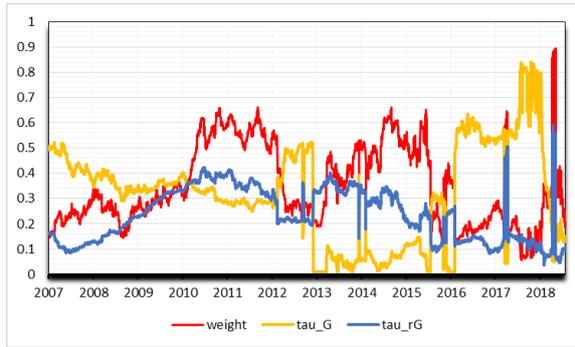
(a) RICIA-RICIM

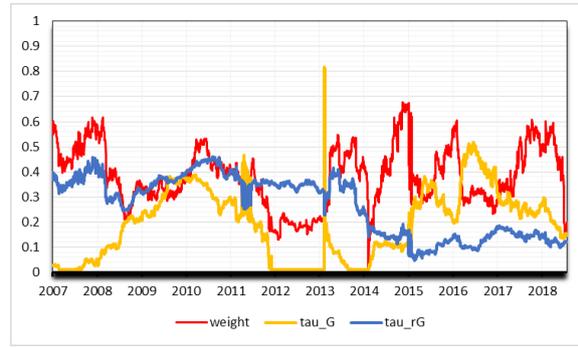
(b) RICIA-RICIE

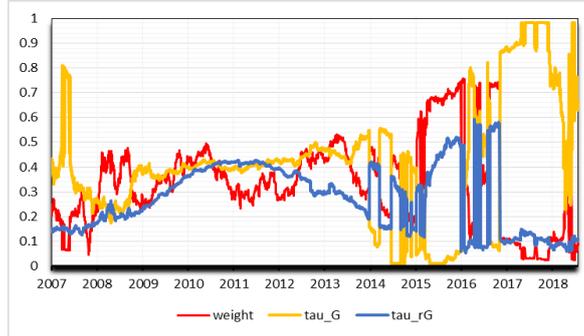
(c) RICIM-RICIE

**Figure 3. Rolling window analysis for the best mixture copula using tau transformation**

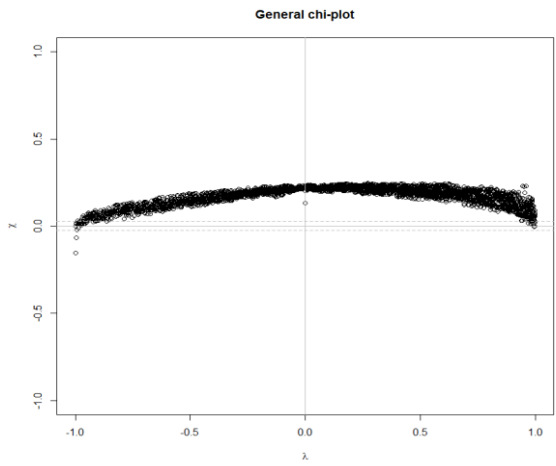
(a) RICIA-RICIM

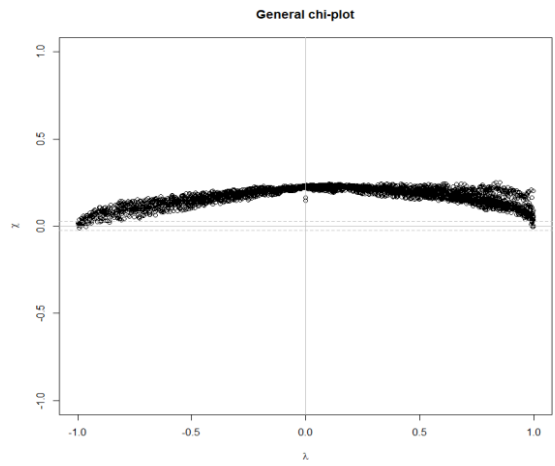
(b) RICIA-RICIE

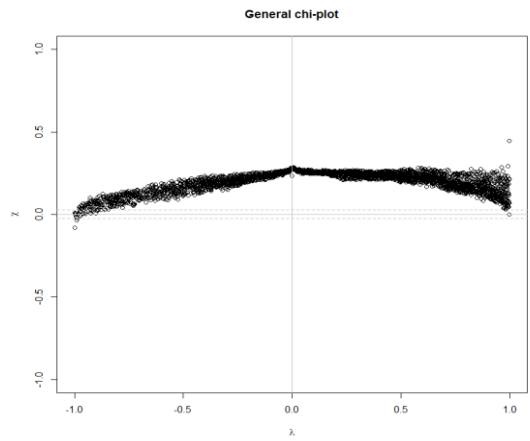
(c) RICIM-RICIE

**Figure 4. Chi-plots illustrated by each pair of commodity indexes**

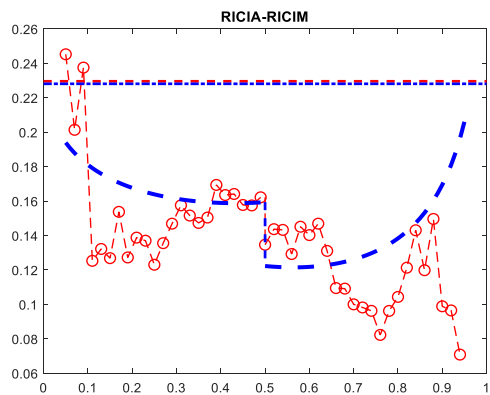
(a) RICIA-RICIM

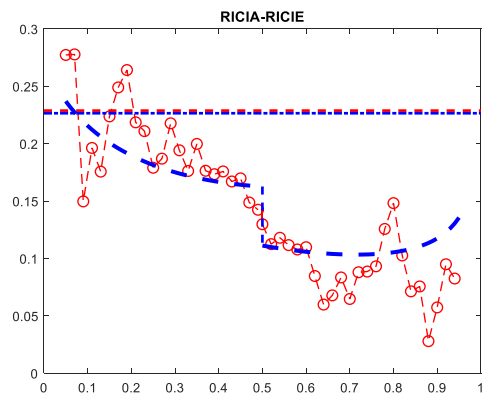
(b) RICIA-RICIE

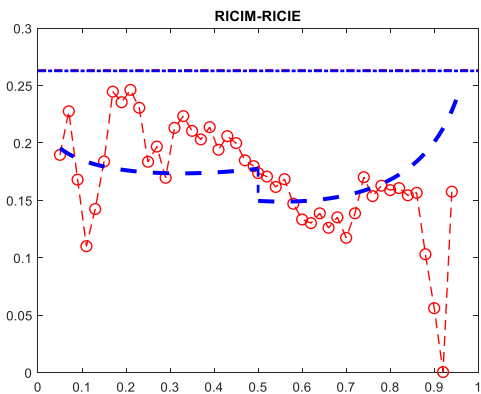
(c) RICIM-RICIE

**Figure 5. Lower-lower and upper-upper quantiles global and local Kendall's tau plots along the main diagonal**

(Note: The red and blue dotted lines represent the empirical and theoretical global Kendall's tau (there are no significant differences between the empirical and the theoretical global Kendall's tau). The blue smooth and the red non-smooth lines with circles represent the theoretical and the empirical local Kendall's tau along the main diagonal.)

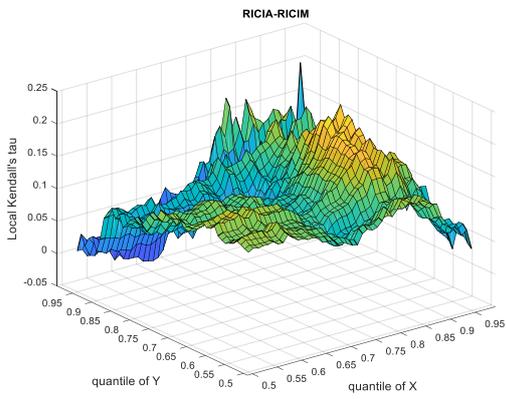
(a) RICIA-RICIM upper tails

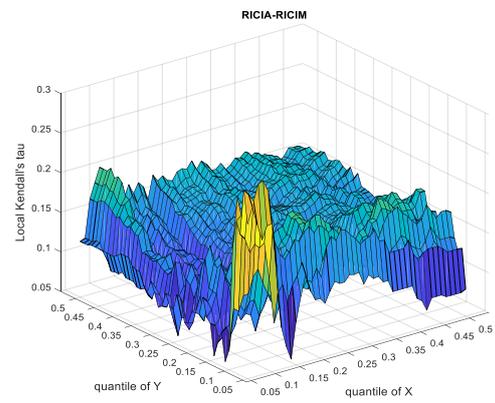
(b) RICIA-RICIM lower tails

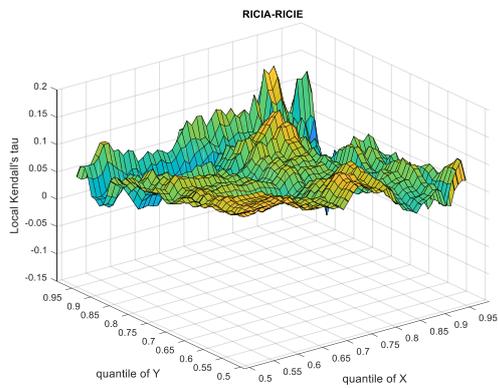
c) RICIA-RICIE upper tails

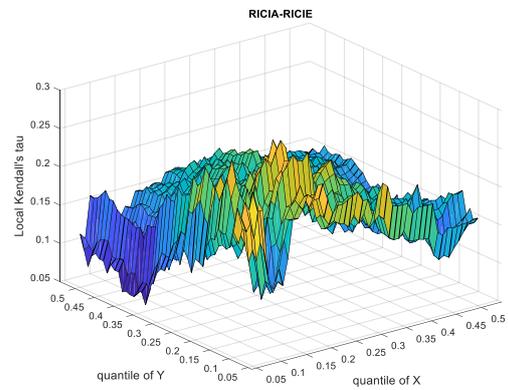
(d) RICIA-RICIE lower tails

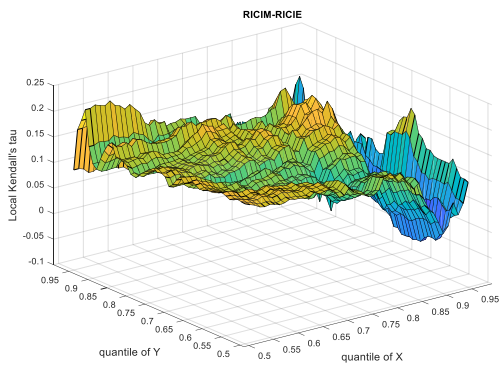
(e) RICIM-RICIE upper tails

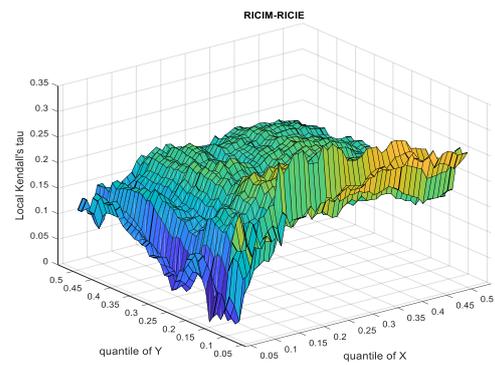
(f) RICIM-RICIE lower tails

**Figure 6. Local Kendall's tau plots for upper and lower tails in quantiles**

(Note: The blue surfaces and yellow surfaces are the theoretical and empirical local dependence surfaces.)

**Tables**

**Table 1. Summary statistics**

|                     | dlnRICIA     | dlnRICIE      | dlnRICIM     |
|---------------------|--------------|---------------|--------------|
| Mean                | 0.002        | 0.009         | 0.143        |
| Median              | -0.020       | 0.048         | 0.040        |
| Maximum             | 11.23        | 29.00         | 19.88        |
| Minimum             | -10.28       | -32.31        | -20.94       |
| Std. Dev.           | 1.056        | 2.191         | 1.410        |
| Skewness            | -0.131       | -0.342        | -0.349       |
| Kurtosis            | 12.14        | 40.72         | 31.52        |
| Jarque-Bera         | 11901.4***   | 202564.0***   | 115840.1***  |
| Stationary analysis |              |               |              |
| ADF                 | -58.85***    | -67.44***     | -65.19***    |
| PP                  | -58.86***    | -67.45***     | -65.16***    |
| KPSS                | 0.093        | 0.109         | 0.336        |
| Observations        | 3414         | 3414          | 3414         |

Note: ***, ** and * denote significance at the 1%, 5% and 10% levels, respectively.

**Table 2. ARMA (1, 0)-GARCH (2, 1) marginal model**

|  | RICIA | | | RICIM | | | RICIE | | |
|---|---|---|---|---|---|---|---|---|---|
|  | Coeff. | Std.E | t-value | Coeff. | Std.E | t-value | Coeff. | Std.E | t-value |
| Cst (M) | -0.014 | 0.014 | -1.007 | 0.034** | 0.016 | 2.079 | 0.034 | 0.023 | 1.441 |
| AR (1) | 0.017 | 0.017 | 1.052 | -0.051*** | 0.017 | -3.001 | -0.045** | 0.017 | -2.546 |
| Cst (V) | 0.006** | 0.003 | 2.201 | 0.020** | 0.008 | 2.325 | 0.018* | 0.010 | 1.690 |
| ARCH ($\alpha_1$) | 0.070*** | 0.023 | 3.065 | 0.081*** | 0.024 | 3.298 | 0.110*** | 0.026 | 4.144 |
| ARCH ($\alpha_2$) | -0.026 | 0.026 | -1.008 | -0.011 | 0.028 | -0.417 | -0.052* | 0.028 | -1.815 |
| GARCH ($\beta_1$) | 0.949*** | 0.014 | 67.79 | 0.920*** | 0.020 | 44.73 | 0.939*** | 0.016 | 56.94 |
| Student (DF) | 6.749*** | 0.955 | 7.066 | 5.924*** | 0.719 | 8.236 | 6.782*** | 0.914 | 7.419 |
| Log Likelihood | -4553.7 | | | -5375.0 | | | -6712.4 | | |
| Akaike | 2.671 | | | 3.152 | | | 3.936 | | |
| Q (10) | 5.173 | | | 11.03 | | | 7.686 | | |
| p-value | [0.818] | | | [0.273] | | | [0.566] | | |
| $Q^2$ (10) | 3.263 | | | 5.936 | | | 6.924 | | |
| p-value | [0.859] | | | [0.547] | | | [0.436] | | |
| ARCH | 1.384 | | | 2.488 | | | 2.179 | | |
| p-value | [0.250] | | | [0.083] | | | [0.113] | | |

Notes: (i) ***, ** and * denote significance at the 1%, 5% and 10% levels, respectively; (ii) Q (10) is the Ljung–Box statistic for serial correlation in the model residuals computed with 10 lags.

**Table 3. Individual copula results**

|  |  | RICIA-RICIM | RICIA-RICIE | RICIM-RICIE |  |  | RICIA-RICIM | RICIA-RICIE | RICIM-RICIE |
|---|---|---|---|---|---|---|---|---|---|
| Gumbel | $C_g$ | 1.262 | 1.256 | 1.324 | Rotated Gumbel | $C_{rg}$ | 1.276 | 1.275 | 1.335 |
|  | s.e. | 0.016 | 0.016 | 0.017 |  | s.e. | 0.016 | 0.016 | 0.017 |
|  | LR | 198.6 | 192.5 | 280.0 |  | LR | 230.4 | 231.1 | 307.8 |
|  | AIC | -395.3 | -383.1 | -558.1 |  | AIC | **-458.8** | **-460.3** | **-613.6** |
|  | BIC | -389.2 | -377.0 | -552.0 |  | BIC | **-452.7** | **-454.2** | **-607.4** |
| Clayton | $C_{CL}$ | 0.463 | 0.463 | 0.548 | Rotated Clayton | $C_{rCL}$ | 0.404 | 0.399 | 0.502 |
|  | s.e. | 0.026 | 0.026 | 0.027 |  | s.e. | 0.026 | 0.025 | 0.027 |
|  | LR | 203.2 | 200.6 | 263.5 |  | LR | 154.2 | 155.1 | 220.7 |
|  | AIC | -404.5 | -399.3 | -525.0 |  | AIC | -306.4 | -308.3 | -439.4 |
|  | BIC | -398.4 | -393.1 | -518.9 |  | BIC | -300.3 | -302.2 | -433.3 |

Notes: (i) *** denotes significance at the 1% level; (ii) bold values indicate the best copula model among all the analysed individual copula models; (iii) s.e. means standard errors; (iv) 3,413 observations.

**Table 4. Mixture copula results**

|  |  | RICIA-RICIM | RICIA-RICIE | RICIM-RICIE |  |  | RICIA-RICIM | RICIA-RICIE | RICIM-RICIE |
|---|---|---|---|---|---|---|---|---|---|
| $C_{mix1}$ |  |  |  |  | $C_{mix2}$ |  |  |  |  |
| weight | $\varphi$ | 0.386 | 0.580 | 0.425 | weight | $\varphi$ | 0.820 | 0.732 | 0.779 |
|  | s.e. | 0.094 | 0.064 | 0.069 |  | s.e. | 0.052 | 0.140 | 0.047 |
| Gumbel | $C_G$ | 1.454 | 1.208 | 1.575 | Rotated Gumbel | $C_{rG}$ | 1.247 | 1.303 | 1.298 |
|  | s.e. | 0.171 | 0.031 | 0.143 |  | s.e. | 0.022 | 0.081 | 0.025 |
| Clayton | $C_{CL}$ | 0.424 | 0.854 | 0.462 | Rotated Clayton | $C_{rCL}$ | 1.143 | 0.532 | 1.221 |
|  | s.e. | 0.099 | 0.150 | 0.083 |  | s.e. | 0.357 | 0.378 | 0.272 |
|  | LR | 242.6 | 242.3 | 326.3 |  | LR | 244.1 | 243.9 | 330.4 |
|  | AIC | -479.3 | -478.7 | -646.6 |  | AIC | -482.2 | -481.8 | -654.8 |
|  | BIC | -460.9 | -460.3 | -628.2 |  | BIC | -463.8 | -463.4 | -636.4 |
| $C_{mix3}$ |  |  |  |  | $C_{mix4}$ |  |  |  |  |
| weight | $\varphi$ | 0.248 | 0.403 | 0.291 | weight | $\varphi$ | 0.677 | 0.505 | 0.638 |
|  | s.e. | 0.064 | 0.101 | 0.058 |  | s.e. | 0.058 | 0.067 | 0.051 |
| Gumbel | $C_G$ | 1.607 | 1.189 | 1.726 | Clayton | $C_{CL}$ | 0.485 | 0.768 | 0.572 |
|  | s.e. | 0.178 | 0.068 | 0.161 |  | s.e. | 0.056 | 0.126 | 0.060 |
| Rotated Gumbel | $C_{rG}$ | 1.220 | 1.375 | 1.251 | Rotated Clayton | $C_{rCL}$ | 0.808 | 0.408 | 0.939 |
|  | s.e. | 0.027 | 0.088 | 0.029 |  | s.e. | 0.180 | 0.073 | 0.165 |
|  | LR | **245.5** | 244.1 | 332.3 |  | LR | 241.1 | 242.0 | 323.7 |
|  | AIC | **-485.1** | -482.3 | **-658.7** |  | AIC | -476.3 | -478.0 | -641.4 |
|  | BIC | **-466.7** | -463.9 | **-640.3** |  | BIC | -457.9 | -459.6 | -623.0 |

Notes: (i) *** denotes significance at the 1% level; (ii) bold values indicate the best copula model among all the analysed mixture copula models; (iii) s.e. means standard errors; (iv) 3,413 observations.

**Table 5. DCC model conditional correlations summary statistics**

|  | Mean | St Dev | Min | Max |
|---|---|---|---|---|
| RICIA/RICIM | 0.414 | 0.128 | 0.094 | 0.740 |
| RICIA/RICIE | 0.414 | 0.127 | 0.180 | 0.772 |
| RICIM/RICIA | 0.469 | 0.133 | 0.130 | 0.732 |

Note: Summary statistics for time-varying conditional correlations $\rho$.

**Table 6. Hedge ratio (long/short) summary statistics**

|  | Mean | St Dev | Min | Max |
|---|---|---|---|---|
| RICIA/RICIM | 0.420 | 4.540 | 0.030 | 224.2 |
| RICIA/RICIE | 0.260 | 0.070 | 0.110 | 0.550 |
| RICIM/RICIA | 0.540 | 0.190 | 0.000 | 1.280 |
| RICIM/RICIE | 0.390 | 0.130 | 0.000 | 0.980 |
| RICIE/RICIA | 0.670 | 0.260 | 0.150 | 1.500 |
| RICIE/RICIM | 0.640 | 2.550 | 0.140 | 126.3 |

Note: Summary statistics for hedge ratios β.

**Table 7. Portfolio weights summary statistics**

|  | Mean | St Dev | Min | Max |
|---|---|---|---|---|
| RICIA/RICIM | 0.710 | 0.140 | 0.000 | 1.000 |
| RICIA/RICIE | 0.840 | 0.120 | 0.180 | 1.000 |
| RICIM/RICIE | 0.690 | 0.170 | 0.010 | 1.000 |

Note: Summary statistics for portfolio weights w.